\documentclass[aps,prl]{revtex4}\usepackage{epsfig}
\usepackage{graphics}
\usepackage{soul}
\usepackage{amsmath}
\begin{document}
\title{Criteria for single photon sources with variable nonclassicality threshold}
\author{Luk\' a\v s Lachman}
\thanks{lachman@optics.upol.cz}
\author{Radim Filip}
\affiliation{Department of Optics, Faculty of Science, Palack\' y University,\\
17. listopadu 1192/12,  771~46 Olomouc, \\ Czech Republic}
\begin{abstract}
Single photon sources are necessary for optical quantum technology. The nonclassicality of emitted single photons manifests itself in diverse experiments and applications. The applications already require different nonclassical aspects of single photons, however, a suitable hierarchy of criteria is missing. We propose variable experimental tests, based on adjustable linear optical networks and single photon detectors, giving a hierarchy of such nonclassicality criteria. The hierarchy goes beyond Hanbury - Brown - Twiss test of photon antibunching and allows us to compare faithfully quality and performance of single-photon sources via their nonclassical properties.
\end{abstract}
\pacs{...}
\maketitle

\flushbottom
\maketitle

\thispagestyle{empty}

\section{Introduction}

Dynamically boosting quantum technology requires rapid development of single photon sources. Such sources are needed in many applications of quantum optics \citep{Eisman,Thew,BB84,Kimble,simulation,Lodahl}. There are many sources producing nonclassical single photon states and their quality is increasing. This nonclassicality is typically witnessed in Hanbury - Brown - Twiss (HBT) test of photon antibunching \cite{mandel, grangier} by a small value of the photon correlation function $g^{(2)}(0)$. It conclusively demonstrates a nonclassical aspect of a single photon state in only one experimental setup. Photon antibunching of single photon states is necessary for some important applications, for example, security of quantum key distribution with a single photon state \citep{BB84,Ekert,Dusek} or quantum metrology with a constraint on a maximal number of photons \citep{Burnett}. For high-quality single photon sources, $g^{(2)}(0)$ is already small \citep{singlePhoton,lodahlC,lodahlSP,senellart,goetzinger} and therefore such HBT test looks always satisfactory and it is not informative anymore.
However, the HBT test uses only single measurement layout among many where the nonclassicality of single photon states can manifest itself. Probably, many other linear optical schemes with single photon detectors can manifest single photon nonclassicality differently to that HBT test \citep{revNoncl, Agarwal,Walmsley,Sperling}.

Such diverse library of nonclassility tests can operationally compare single photon sources according to their sensitivity to a sequence of criteria.  One from two sources can provably generate light manifesting the nonclassicality in a test whereas the other cannot. Due to our previous development in ab initio derivation of nonclassicality criteria \citep{filipLach}, we can derive criteria for any given measurement layout. Such criteria are operational because they uniquely correspond to that layout. It is still possible that in a different measurement layout nonclassicality of light from single photon sources will not manifest despite doing so for ideal single photon states without the multiphoton contribution. It will therefore be fruitful to have such direct measurement schemes giving sensitive criteria allowing classification of single photon states beyond HBT measurement. Such operational criteria, with a more demanding nonclassicality threshold for the sources, will simultaneously give a new picture about a variety of nonclassical aspects with a single photon state beyond essential photon antibunching.

\begin{figure}[t]
\centerline {\includegraphics[width=0.6\linewidth]{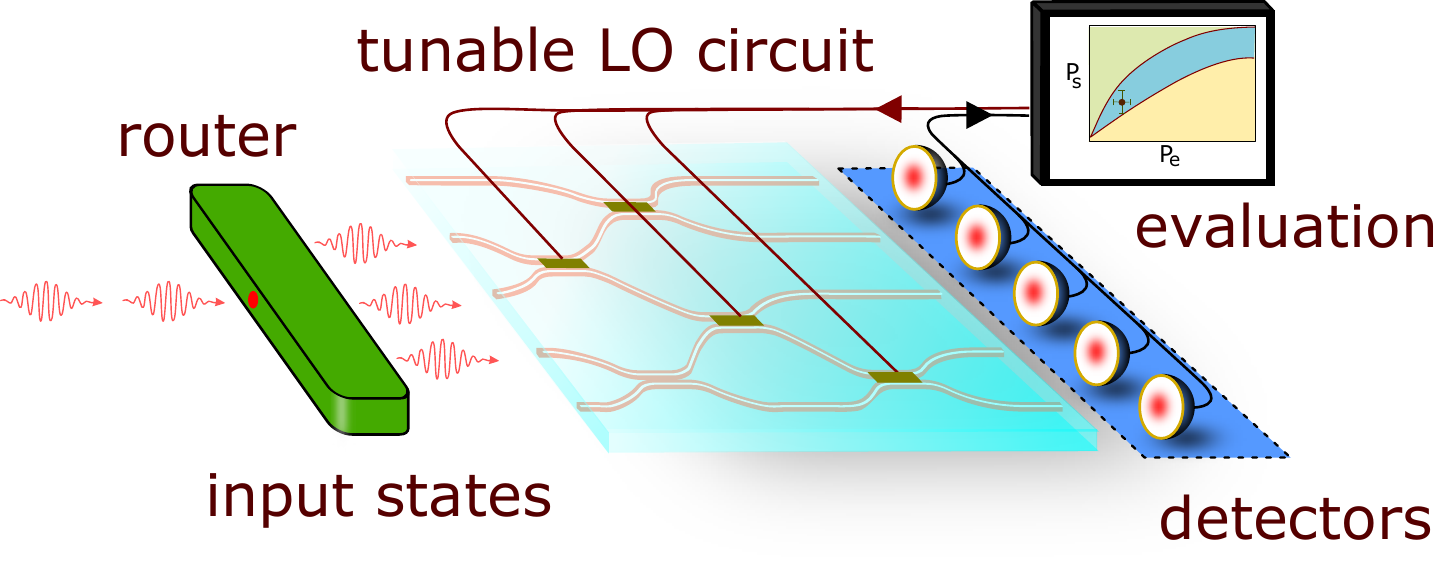}}
\caption{Measurement layout for nonclassicality criteria with a variable threshold based on tunable integrated optics. A single photon periodically comes to a router that produces $N$ copies of the state by delaying some of the inputs. These states are further guided through a linear optical (LO) circuit towards $K$ detectors. The parameters of the LO circuit can be driven electrically to manipulate the nonclassicality threshold for the single photon source. The success and error probabilities of detection events are evaluated and compared in a computer with thresholds of nonclassicality derived for a specific settings of the LO circuit.}
\label{idea}
\end{figure}

For these reasons, we propose both a methodology and relevant examples of criteria with arbitrarily variable nonclassicality thresholds for single photon states. We also extend the proposal to multi-copy criteria exploiting interference among single photon states. Since all these criteria incorporate interference, they go beyond HBT measurement. Further, they can impose an arbitrarily variable condition on single photon states and therefore they can establish a hierarchy of measurement layouts for single photon sources.  These useful layouts single out particular linear optical circuits, where the manifestation of nonclassicality requires more profound suppression of multiphoton contribution. They also open a new insight into nonclassical aspects of single photon states.

\section{Detection of nonclassicality}
\subsection{Measurement layout}
A general platform used for such measurement on light from single photon sources needs an optical router and an adjustable K$\times$K linear optical (LO) circuit where $N < K$ initial states of light can interfere, as is depicted in Fig.~\ref{idea}. All the LO interference experiments can be already integrated to an optical chip and together with integrated detectors they can form variable detection units for single photon sources. At an output, photons are non-trivially split among all $K$ modes and are detected by conventional single-photon avalanche photo-diodes (SPADs). If $N$ ideal single photons without multiphoton contributions are injected to different input ports, then maximally $N$ ideal SPADs can produce a signal (a click). If more SPADs than $N$ give a click, it is either because of a multiphoton noise in the input states or dark count events. For the $N$ inputs with optical signal, we compare directly $N$ simultaneous clicks of the SPADs, which are considered as {\em successful} events for single photon states, with multiphoton {\em errors} corresponding to $N+1$ simultaneous clicks. As the SPADs are insensitive to photon numbers, we model their response using positive-operator-value-measure with the components $\Pi_0=|0\rangle\langle 0|$ (no click) and $\Pi_{1+}=1-|0\rangle\langle 0|$ (click). A detector efficiency can be characterised and included to the description of a linear multi-port. This description is prior knowledge used to derive the criteria. On the other hand, dark counts in the SPADs producing false detection events are accounted as an additional background noise of an inspected state. They can only make results worse. This treatment of detector efficiency and dark counts saves the following criteria from systematic errors, as in previous works \citep{nonG1, nonGN, Obsil}. The interferometric network is {\em adjustable}; all beam splitters and the phase shifts between them can be manipulated to reach the desired layout flexibly  \cite{tuning1,tuning2}. Adjustable optical circuits can be implemented using integrated optical technology \citep{integratedOptics,integrateOptics2,integrateOptics3}. Together with upcoming integrated SPADs, it can form a versatile miniaturized detector for new tests of the quality of single photon sources.

\subsection{Ab initio nonclassicality witness}
Nonclassicality of light manifested in each network is defined as an incompatibility of the detection events with the classical coherence theory of light \cite{glauber}. Therefore, the nonclassicality refers to incompatibility with a convex set of multimode states
\begin{equation}\label{classical}
\rho_{cl}=\sum_{\omega}\int P_{\omega}(\alpha)\vert \alpha \rangle_{\omega} \langle \alpha \vert_{\omega} \mathrm{d}^2 \alpha,
\end{equation}
where $\vert \alpha \rangle_{\omega}$ is a coherent state having mode label $\omega$ and $P_{\omega}(\alpha)$ satisfies all the requirements for a probability density function. Considering all multimode states with different $\omega$, any detection of nonclassicality excludes all classical states with any level of classical (first-order) coherence. To derive these criteria, we consider a linear combination of the probabilities
\begin{equation}
W_a(\rho)=P_s+a P_e,
\label{W}
\end{equation}
where $P_s$ is the probability of success ($N$ from $K$ detectors click for $N$ states at the input), $P_e$ is the probability of error (more than $N$ clicks for $N$ states at the input) and $a$ is a free parameter. Nonclassicality manifests when $W_a(\rho)$ exceeds a value $W_{max}(a)$ covering all mixtures of multimode coherent states (\ref{classical}). The linearity of the witness $W_a(\rho)$ guarantees that $W_{\max}(a)$ is determined by optimizing over pure coherent states occupying only a single mode. The amplitude $\alpha_{opt}(a)$ of the optimal single-mode state depends on the parameter $a$. It is a complex task to derive criteria for all possible LO multi-ports in the Fig.~\ref{idea}. Many of complex layouts can be solved only numerically. We restrict the presentation here to a detailed description of the most relevant and simplest variable criteria: (i) a variable Mach-Zehnder interferometer using single photon interference and (ii) a variable network with interference between two single photon states. A more systematic analysis of more complex layouts, straightforward but too technical, is left for a further investigation.

\begin{figure}[t]
\centerline {\includegraphics[width=0.9\linewidth]{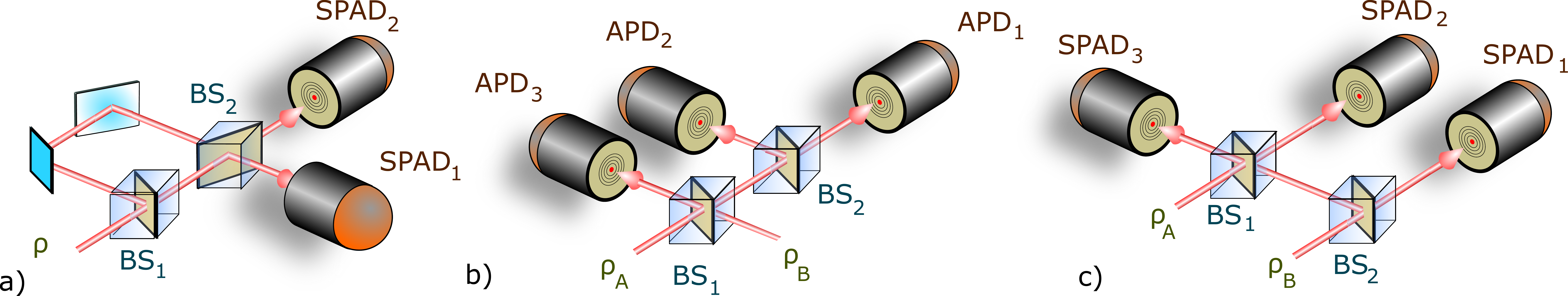}}
\caption{Examples of the simplest layouts for variable nonclassicality criteria beyond Hanbury-Brown-Twiss measurement. a) Mach-Zehnder interferometer with adjustable transmissions of $\mbox{BS}_2$ provides a hierarchy of the criteria based on a single-photon interference effect at BS$_2$. A success event happens when SPAD$_1$ registers a click. An error event appears when both detectors click. b) Extended Hong-Ou-Mandel interferometer is exploited for nonclassical tests on two states $\rho_A$ and $\rho_B$. A success occurs when SPAD$_1$ and SPAD$_2$ click simultaneously and error is detected when all three detectors register signal. c) Layout giving the hierarchy of nonclassical conditions for two states $\rho_A$ and $\rho_B$. A success event happens when SPAD$_1$ and SPAD$_2$ register a click simultaneously. An error event appears when all three SPADs click. Both BS$_1$ and BS$_2$ are tuned to vary nonclassicality criteria.}
\label{setups}
\end{figure}

\subsection{Realistic single photon states}
We cannot realize the applicability of the schemes in Fig.~\ref{idea} to derive new criteria, without testing them on a typical class of experimentally relevant single photon states. High-quality sources of single photon states currently approach a density matrix $\rho_{\eta}\otimes\rho_{\bar{n}}$, where $\rho_{\eta}=\eta \vert 1 \rangle \langle 1 \vert+(1-\eta)\vert 0 \rangle \langle 0 \vert$ is an ideal single photon without multiphoton contributions with $\eta$ standing for the emission and collection efficiency. The deteriorating background noise obeys Poissonian statistics $\rho_{\bar{n}}=e^{-\bar{n}}\sum_{n=0}^{\infty}\frac{\bar{n}^n}{n!}\vert n \rangle \langle n \vert$ with mean number of photons $\bar{n}$. Non-zero collection efficiency $\eta$ and finite amount of Poissonian background noise {\em never} disrupt the detection of nonclassicality if the HBT measurement is used. It implies from the maximization of (\ref{W}) applied on the HBT measurement. The procedure leads to $P_s>\sqrt{P_e}$ \cite{filipLach}, where $P_s$ is a probability of a single detection event on one SPAD (irrespectively to the other) and $P_e$ is a probability of a coincidence event, when both SPAD register photons simultaneously. This is exactly the criterion used in practice by experimentalists to verify nonclassicality by HBT measurement. Notably, it is derived without any use of normally-ordered correlation functions forming $g^{(2)}(0)$. It only requires a prior knowledge about a detection layout and probabilities of detector clicks. This ab initio approach is flexible and allows derivation of new criteria for any measurement layout. We will therefore use it to derive the variable nonclassicality criteria and focus mainly on the experimentally very relevant case of high-quality single photon states $\rho_{\eta}\otimes\rho_{\bar{n}}$ with low background noise.

\subsection{Single-copy variable nonclassicality criteria}
A natural feasible extension of the previous layout for nonclassical detection is an unbalanced version with a variable beam splitter (BS). In the limit of states with small multiphoton contribution, the nonclassicality criterion turns to be
\begin{equation}
P_s>\sqrt{\frac{T}{1-T}}P_e^{1/2},
\label{HBT}
\end{equation}
where $T$ is the transmission of the BS \cite{filipLach}. Although adjusting $T\in (0,1)$ can alter arbitrarily the nonclassicality threshold for $P_s$, this criterion (\ref{HBT}) is actually not more demanding than the HBT test for the relevant states $\rho_{\eta}\otimes\rho_{\bar{n}}$. It is simple to verify that the probabilities approximated by $P_s\approx (\eta+\bar{n})T$ and $P_e \approx 2T(1-T)\bar{n}\eta$ in condition (\ref{HBT}) reveal that states with $\eta>0$ radiated together with arbitrary $\bar{n}>0$ of background noise are always nonclassical for every transmission $T\in (0,1)$. We have checked that it is also true for a similar extension of detection schemes with three and four detectors presented in Ref~\cite{filipLach}, that are utilized to detect nonclassicality of states $(\rho_{\eta}\otimes \rho_{\bar{n}})^{\otimes 2}$ and $(\rho_{\eta}\otimes \rho_{\bar{n}})^{\otimes 3}$ injected to single input port. Therefore, it indicates a conjecture that criteria which impose a variable nonclassicality condition for the relevant states require interfering networks.

The Mach-Zehnder interferometer based on first-order coherence is the simplest choice. However, it is not sufficient to combine sequentially the interferometer and HBT measurement, i. e. using only the HBT scheme in one of the outputs of the Mach-Zehnder interferometer. For any interferometer with different $T_1$, $T_2$ and a relative phase $\phi_{MZ}$, a state emerging in one of the output modes of the interferometer has still the form $\rho_{\eta}\otimes\rho_{\bar{n}}$, and therefore, a more demanding nonclassicality condition cannot be also derived for this case. To construct more demanding criteria than HBT criteria, we have to {\em combine} the interferometer with a two-detector test as is depicted in Fig.~\ref{setups}a). This layout acts efficiently as an adjustable BS only for monochromatic light \citep{tuning1}. If the noise $\rho_{\bar{n}}$ experiences a different $\phi_{MZ}$ than the state $\rho_{\eta}$, the nonclassicality manifests itself differently compared to HBT test. This happens when the noise $\rho_{\bar{n}}$ behaves as polychromatic in Mach-Zehnder interferometer. There are two paths by which a state of light propagates towards $\mbox{SPAD}_1$ and $\mbox{SPAD}_2$. Ideal interference of coherent states in both paths results in a suppression of clicks registered by one of the SPAD and an increase in the number of clicks of the second SPAD. Setting an adequate relative phase $\phi_{MZ}$ between both paths can cause that one SPAD always clicks and the other never even when a classical monochromatic state propagates through the interferometer. However, this behavior requires that the transmissions of BSs in the interferometer satisfy $T_1+T_2=1$. If it is not the case, the coherent state is split in the interferometer and coincidence clicks of both SPADs can be detected in this layout. Complete suppression of these events is possible only when single photon states, even polychromatic, are guided through the interferometer and therefore it is beyond classical optics.

\begin{figure*}[t]
\centerline {\includegraphics[width=0.95\linewidth]{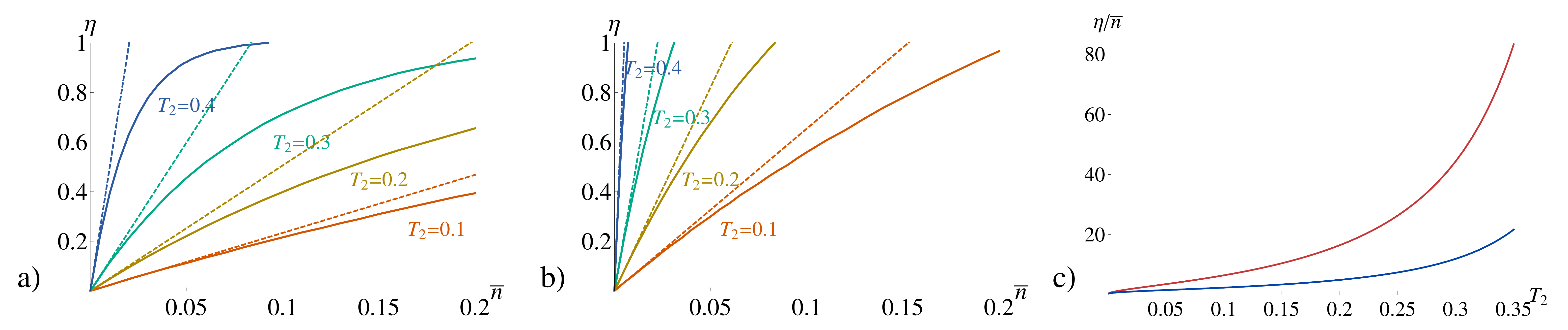}}
\caption{Hierarchy of thresholds of nonclassicality based on Mach-Zehnder interferometer in Fig.~\ref{setups}a) for $T_1=0.5$ and variable transmittance $T_2$. The inspected state has a form $\rho_{\eta}\otimes \rho_{\bar{n}}$. Its single photon component $\rho_{\eta}$ behaves coherently in the interferometer \textit{(a)} or incoherently \textit{(b)}. The solid lines represent the numerical results and the dashed ones correspond to linear approximations. Figure \textit{(c)} demonstrates the difference of approximate thresholds on the ratio $\eta/\bar{n}$ between coherent (blue) and incoherent (red) limits.}
\label{machZ}
\end{figure*}

Let us consider a successful event as a click of the SPAD with a higher click probability and denote the probability by $P_s$. An error corresponds to a coincidence of the clicks at both SPADs and the probability of error is designated by $P_e$. The threshold is derived from maximizing function (\ref{W}) over all classical states. It was verified that the optimum is reached for monochromatic coherent light. The relative phase $\phi_{MZ}=2\pi \omega d/c$ between coherent states choosing different paths is sensitive to the frequency $\omega$ and to the path difference $d$. The constant $c$ denotes the speed of light. Thus, the relative phase is a product of the fixed parameter $d$ given by the layout and of the frequency $\omega$ which is determined from optimizing function (\ref{W}). Whereas the optimal amplitude of the coherent state depends on a choice of the free parameter $a$, the optimal frequency $\omega$ is such that the relative phase obtains $\phi_{MZ}=2\pi n$ for arbitrary $a$, where $n$ is an integer. Since the optimum is independent on $d$, a single nonclassicality threshold can be exploited for a layout with any path difference $d$. Therefore the path difference can be adjusted so that single photon states  $\rho_{\eta}$ and noise $\rho_{\bar{n}}$ behave differently in the interferometer. If the interferometer is unstable and $d$ fluctuates, the threshold is always decreased. It means that thresholds derived with an assumption that the unknown parameter $d$ is fixed are reliable even for unstable interferometers. The nonclassicality threshold for $P_s$ and $P_e$ can not be expressed analytically. However, it can be straightforwardly calculated numerically. The criteria are variable by changing  $T_2$ for a fixed $T_1$. The numerical solution for $T_1=1/2$ and several values of $T_2$ is depicted in Fig.~\ref{machZ} a). For high quality states with small multiphoton contribution one can find a useful approximate formula
\begin{equation}
P_s>f(T_1,T_2)P_e^{1/2}
\label{MZ}
\end{equation}
where $f(T_1,T_2)$ is a function of BS’s parameters $T_1$ and $T_2$. Let us denote $\Delta=T_1+T_2-1$ and fix $T_1=T$. For setups with non-vanishing but small $\vert \Delta \vert \ll 1$ the function obtains
\begin{equation}
f(T,\Delta)\approx \frac{2\sqrt{T(1-T)}}{\vert \Delta \vert}.
\end{equation}
The nominator is fixed and the criteria change due to the variable denominator. 
It is used to find analytic approximations of the criteria.

We analyzed both extremes of monochromatic (coherent) and polychromatic (incoherent) limits of $\rho_{\eta}$ to test if the criteria (\ref{MZ}) can give really gradually varying nonclassicality criteria for the relevant states $\rho_{\eta}\otimes\rho_{\bar{n}}$. Monochromatic single photons perfectly interfere at the BS$_2$ whereas polychromatic do not at all.
If the monochromatic state $\rho_{\eta}$ propagates through a stable interferometer, a click is registered on $\mbox{SPAD}_{1,2}$ with probabilities
\begin{eqnarray}
P_{s,1}&=&\eta\left[ T_1 R_2+T_2 R_1+2 \cos \phi_{MZ}\sqrt{T_1 T_2 R_1 R_2}\right]\nonumber \\
P_{s,2}&=&\eta\left[ T_1 T_2+R_1 R_2-2 \cos \phi_{MZ}\sqrt{T_1 T_2 R_1 R_2}\right],
\end{eqnarray}
where $\phi_{MZ}$ is the relative phase acquired between the two paths of the interferometer and $R_i$ is the reflectivity of $\mbox{BS}_i$, where $i=1,2$. Since the thresholds of nonclassicality cover coherent states with any relative phase $\phi_{MZ}$, a single criterion can be exploited for states with any $\phi_{MZ}$ acquired in the interferometer. If the state $\rho_{\eta}$ is averaged over more modes at many different frequencies $\omega$, the phase $\phi_{MZ}$ dependent terms gradually vanish, and the probabilities approach the incoherent limit
\begin{equation}
P_{s,1}=\eta ( T_1 R_2+T_2 R_1),P_{s,2}=\eta ( T_1 T_2+R_1 R_2).
\end{equation}
The same response of detectors occurs when the interferometer is unstable.
The model of state $\rho_{\eta}\otimes\rho_{\bar{n}}$ involves error events typically caused by background noise.
Figs.~\ref{machZ} a) and \ref{machZ} b) present the influence of background noise with mean photon number $\bar{n}$ on the detection of nonclassicality for different arrangements of $T_1$ and $T_2$ in both monochromatic and polychromatic limits of the state $\rho_{\eta}$. Fig.~\ref{machZ} c) demonstrates a difference of thresholds on nonclassicality between these limits. Recall, that for all these states nonclassicality is always detected using HBT (\ref{HBT}). Apparently, for both ideally monochromatic and also for polychromatic states, it is gradually more demanding to prove their nonclasical features for the same states with criterion (\ref{MZ}). Visibly, the criteria (\ref{MZ}) give a variable nonclassicality threshold for both limits of monochromatic and polychromatic states $\rho_{\eta}\otimes\rho_{\bar{n}}$ even for arbitrary small $\eta$ and $\bar{n}$. The same variability also happens for an intermediate case of partially monochromatic (coherent) light.

The criteria are variable by changing $T_2$ for a fixed $T_1$. The thresholds can be parameterised by a ratio $\eta/\bar{n}$ in a practical region of high quality single photon states. In this model, the factor $\eta/\bar{n}$ determines a ratio of single photon probabilities from the ideal source and the background noise. This ratio can be estimated from HBT measurement in this region of high quality single photon states employing approximations $P_s \approx (\eta+\bar{n})/2$ and $P_e \approx \eta \bar{n}/2$. However, the operational meaning of the ratio relates it to nonclassicality detection presented here. Mach-Zehnder interferometer is a first known example of how the layout can vary the nonclassicality threshold in a way useful for single photon sources. The hierarchy can be controlled by altering transmission $T_2$ for arbitrarily fixed $T_1$. Setting $T_2=0$ or $T_2=1$ reduces the layout to an examination of criterion (\ref{HBT}) and therefore the condition tolerates arbitrary noise. In contrast, reaching $1-T_1-T_2 \approx 0$ results in a very demanding condition on the state. 

The threshold is tunable for these states due to first order coherence of the state $\rho_{\eta}$. Let us assume $\vert \Delta \vert \ll 1$, where $\Delta=T_1+T_2-1$. The state $\rho_{\eta}\otimes \rho_{\bar{n}}$ exhibits nonclassicality in the practical limit of states with small $\eta$ and $\bar{n}$, if
\begin{equation}
\eta>\frac{8 T_1^2(1-T_1)^2}{\Delta^2}C(\rho_{\eta},d)\bar{n},
\label{modelTh}
\end{equation}
where $\bar{n}$ is very small and the factor $C(\rho_{\eta}, d)$  depends on the monochromaticity of the state $\rho_{\eta}$ and on the path difference $d$. The factor for monochromatic $\rho_{\eta}$ or for interferometer with $d=0$ reaches $C=1$. On the other hand, polychromatic limit of the state $\rho_{\eta}$ leads to $C=2/(1-2 T_1+2 T_1^2)$ for any $d$. Thus, the recognition of nonclassicality is dependent on the coherent features of the state $\rho_{\eta}$. Apparently, a class of realistic states $\rho_{\eta}\otimes\rho_{\bar{n}}$ can still manifest its nonclassicality in the  layout (\ref{MZ}) with $\Delta \neq 0$, since an ideal single photon state $\rho_{\eta}$ would always exhibit nonclassicality. Such cases appear for monochromatic single photon states as well as for polychromatic single photon states. If the noise $\rho_{\bar{n}}$ propagates incoherently in the interferometer then adjusting $\Delta$ allows us to recognize some states as better single photon sources, because they surpass more demanding tests.  The variable test may require an experimental realization of the interferometer with a path difference sufficiently large such that the noise propagates incoherently in the interferometer. For these states $\rho_{\eta}\otimes\rho_{\bar{n}}$, the Mach-Zehnder interferometer is a minimal layout to obtain such a variable and arbitrarily demanding hierarchy of nonclassical criteria. It can broadly stimulate experimental teams developing single photon sources that can be further operationally categorised beyond a HBT analysis. It means that we can identify a (linear optical) scheme in which light from one source manifests its nonclassical character better than from other sources.

\subsection{Two-copy variable nonclassicality criteria}
We can now extend the methodology to the two-photon Hong-Ou-Mandel (HOM) interference effect. It is attractive because it can bring a variable test of nonclassicality in a principally different interference scheme. Differently to the layout at Fig.~\ref{setups} a) and previous derivation, it uses two-mode classical states for the derivation of nonclassicality threshold. Motivated to find the simplest layout, we first tried to extend the traditional HOM test for nonclassicality with two detectors \cite{hong,hongCr} to be able to define a successful detection event for a pair of ideal single photon states. It can be tempting to consider a scheme which interferes two copies of emitted states at a beam splitter, splits then one output to three SPADs and discards the other. It is a sequential combination of a photon bunching and nonclassicality criteria in Ref.~\cite{filipLach}. If success is defined as a click of two detectors and error as a click of all three, then for any $\eta>0$ and finite $\bar{n}$ it is always possible to prove nonclassicality for any pair of states $\rho_{\eta}\otimes\rho_{\bar{n}}$. Such a simple layout, therefore, does not give criteria beyond the HBT measurement. 

To go beyond HBT criteria, we extend the HOM interferometer with the three detectors differently. It consists now of two beam splitters $\mbox{BS}_1$ and $\mbox{BS}_2$ and three detectors $\mbox{SPAD}_1$, $\mbox{SPAD}_2$ and $\mbox{SPAD}_3$ detecting all the outputs as depicted in Fig.~\ref{setups} b). $\mbox{BS}_1$ performs two-photon interference of the incoming states, as in the HOM measurement, and $\mbox{BS}_2$ splits one of the outputs toward $\mbox{SPAD}_1$ and $\mbox{SPAD}_2$ to detect pairs of photons. The last detector $\mbox{SPAD}_3$ detects the second output without any splitting. Success corresponds here to a coincidence click of $\mbox{SPAD}_1$, $\mbox{SPAD}_2$ caused by the photon bunching effect irrespective of the results of $\mbox{SPAD}_3$. Error is represented by the simultaneous click of all three detectors, which never happens without a multiphoton contribution. In this layout, classical interference between two coherent states with a stable phase can, however, leads to no photons at $\mbox{SPAD}_3$ (no errors) and still some coincidences at $\mbox{SPAD}_1$ and $\mbox{SPAD}_2$. Such events can cause a systematic failure of the method if such classical interference is not eliminated.
\begin{figure*}[!t]
\centerline {\includegraphics[width=0.95\linewidth]{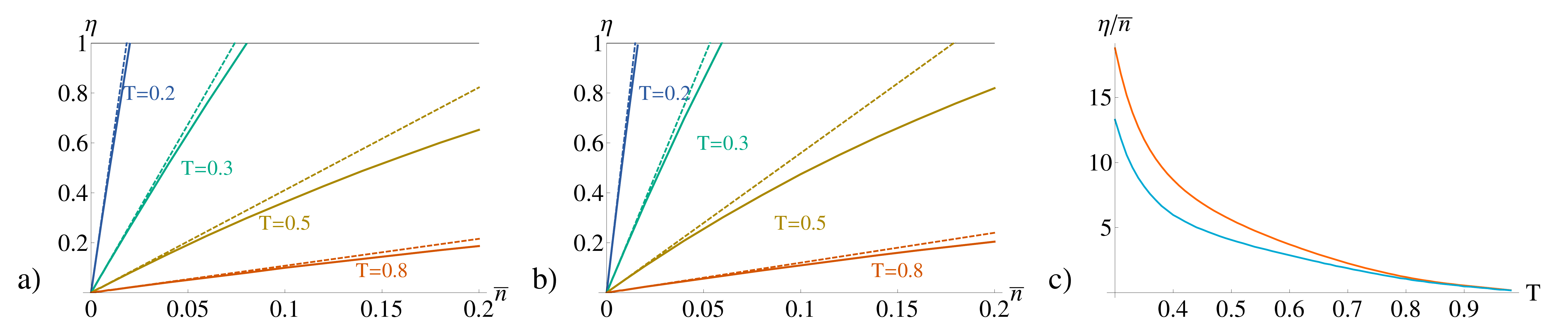}}
\caption{Witnessing nonclassicality of a two-copy state $(\rho_{\eta}\otimes \rho_{\bar{n}})^{\otimes 2}$ derived for the interfering setup depicted in Fig.~\ref{setups}b). The incoming states $\rho_{\eta}$ are indistinguishable (a) or distinguishable (b) at BS$_1$. Whereas the solid lines represent the exact numerical solution, the dashed lines correspond to the linear approximation. (c) The difference between the linear threshold of the distinguishable (red) and indistinguishable (blue) states $\rho_{\eta}$ for different $T$. The sensitivity of the nonclassical threshold to the indistinguishibility brings a new aspect observable in this setup.}
\label{tritter}
\end{figure*}
To eliminate such cases, the measurement can be easily performed for various random phases of both input states. This is a new aspect of this three-detector scheme.  Such nonclassicality criteria will detect the states beyond the phase-randomized multimode classical states with a density matrix
\begin{equation}
\rho_{cl}=\sum_{\omega_a,\omega_b} \int P_{\omega_a,\omega_b}(\vert \alpha \vert ,\vert \beta \vert) \rho_{\omega_a}(\vert \alpha \vert) \otimes \rho_{\omega_b}(\vert \beta \vert) \mathrm{d}\vert \alpha \vert \mathrm{d}\vert \beta \vert,
\label{clState}
\end{equation}
where $a$ and $b$ denote the input spatial modes, indices $\omega_{a,b}$ cover all degree of freedom of the initial modes involving frequency and polarization, $P_{\omega_a,\omega_b}(\vert \alpha \vert ,\vert \beta \vert)$ is a probability density function related to modes $\omega_a$ and $\omega_b$ and $\rho_{\omega}$ is density matrix of a state occupying a collection of background modes $\omega$ and obeying Poissonian statistics
\begin{equation}
\rho_{P,\omega}(\vert \alpha \vert)=e^{-\vert \alpha \vert}\sum_{n=0}^{\infty}\frac{\vert \alpha \vert^n}{n!}\vert n \rangle_{\omega} \langle n \vert.
\label{clSt}
\end{equation}
Due to summation over different mode indices $\omega_a$ and $\omega_b$, any classical phase-randomized multimode light is expressed by (\ref{clState}). Note, two-photon interference depends on the indistinguishability of photons at the BS$_1$. The multi-indices $\omega_a$, $\omega_b$ and $\omega$ therefore describe all mode features of the light beams. Differently to the previous case,  the value $W_{max}(a)$ is now reached when we interfere two coherent phase-randomized states. The optimal amplitudes $\vert \alpha(a) \vert$ and $\vert \beta(a) \vert$ depend on the free parameter $a$ in (\ref{W}). The calculation of $P_s$ and $P_e$ is straightforward, however an optimization is only numerical, because the phase-randomization introduces Bessel functions as shown in appendix. For states with very small multiphoton contributions, the nonclassicality criterion can be approximated as $P_s>f(T_1,T_2) P_{e}^{2/3}$. The function $f(T_1,T_2)$ depends on $T_1$ and $T_2$ which  represent transmission of BS$_1$ and BS$_2$. 

To test usefulness of this criterion, we used two copies $(\rho_{\eta}\otimes\rho_{\bar{n}})^{\otimes 2}$ and analyzed them at both limits of indistinguishable and distinguishable states $\rho_{\eta}$. The background noise $\rho_{\bar{n}}$ never interferes. A witness built for the layout with $T_1=1/2$ imposes the most tolerant condition on $\bar{n}$ for a given $\eta$ independently to a setting of the transmission $T_2$. Although the condition is the most lenient for that layout, it does not tolerate arbitrarily large $\bar{n}$. On the contrary, arbitrary strictness can be achieved when $T_1$ approaches either $T_1=0$ or $T_1=1$. In the extreme cases, the detectors measure a state that is factorized to a state observed with detectors SPAD$_1$ and SPAD$_2$ and to a state registered only by SPAD$_3$. Therefore even complete suppression of the error probability achieved by the ideal state $\rho_{\eta}\otimes \rho_{\eta}$ can be explained classically. Apart from these two extreme values of $T_1$, states $\rho_{\eta}$ without multiphoton contribution will always manifest nonclassicality. The tolerance to background noise contribution can be manipulated with the beam splitter asymmetry, similarly as in the Mach-Zehnder layout. Contrary to that layout, this witnessing is sensitive to the indistinguishibility of the states $\rho_{\eta}\otimes \rho_{\eta}$ and therefore it tests nonclassicality that is not manifested in the Mach-Zehnder interferometer.
However, this setup still does not allow us to gain an arbitrarily tolerant nonclassicality threshold yet and therefore we cannot completely vary the nonclassicality threshold for $\rho_{\eta}\otimes\rho_{\bar{n}}$.

A fully variable nonclassical threshold can be reached by re-arranging the layout as shown in Fig.~\ref{setups} c). The layout apparently does not depend on first-order coherence of the emitted single photon states, only on their indistinguishability. It makes this test complementary to the previous one employing Mach-Zehnder interferometer. A combination of two states interference at BS$_1$ and anti-bunching at BS$_2$ is different than in the previous case. Successful events are defined in this case as a simultaneous clicks in the detectors $\mbox{SPAD}_1$ and $\mbox{SPAD}_2$. Error occurs when all the three detectors click. The success probability quantifies events when both states are transmitted without any two-photon interference. Apparently, the scheme goes also beyond the HOM interference effect although still uses two copies of the emitted state. Classical coherent states can suppress clicks of $\mbox{SPAD}_3$ due to destructive interference on BS$_1$. Therefore, we also consider here the relative phase between both initial states to be random in the measurement. For states with a small multiphoton contribution, the derivation gives an approximate threshold of nonclassicality
\begin{equation}
P_s>f(T_1,T_2) P_{e}^{2/3},
\label{MZHOM}
\end{equation}
where the function $f(T_1,T_2)$ has an extensive analytic expression. The criteria form a hierarchy of operational conditions on a two copy state $(\rho_{\eta}\otimes \rho_{\bar{n}})^{\otimes 2}$ with non-interfering noise $\rho_{\bar{n}}$. Figs.~\ref{tritter} a) and \ref{tritter} b) show these criteria can be manipulated freely even in a practical region of small $\eta$ and $\bar{n}$ for both distinguishable and indistinguishable states. The hierarchy can be smoothly controlled by a parameter $T=T_1=T_2$, as is depicted in Fig. \ref{tritter} c). The strictest threshold is reached by $T$ close to zero. On the other hand, setting $T$ almost one gives very tolerant threshold. In that extreme case, the condition can be approximated by
\begin{equation}
\eta>\sqrt{1-T} \bar{n},
\end{equation}
which holds for both distinguishable and indistinguishable states $\rho_{\eta}$ only if $1-T \ll 1$. Note, that setting $T=1$ implies the beam splitters completely transmit the full signal and therefore the error events never occur for any state. Except for this limited case, the criteria impose an arbitrarily tolerant condition on the noise.

\section{Conclusion and outlook}
Our approach suggests that nonclassicality of light from single photon emitters can be straightforwardly verified beyond standard HBT measurement which is only one, fixed and too easy sufficient condition for nonclassicality. True high quality single photon source has to fulfill many strict nonclassicality conditions. We have suggested a methodology to find such nontrivial sufficient conditions for nonclassicality. Explicitly, the two proposed simplest examples of criteria (\ref{MZ}) and (\ref{MZHOM}) use different setups to manifest different aspects of nonclassicality of light. Both criteria go beyond standard HBT test (\ref{HBT}) because their derivation involves either single photon or two photon interference. The single copy variable criteria exploiting Mach-Zehnder interferometer forms a hierarchy only when the noise $\rho_{\bar{n}}$ has a broader spectrum than the signal $\rho_{\eta}$ or when both signal and noise have low visibility. If it is not the case then the hierarchy can be built only for tests with two copies tests. These operational hierarchies are formulated in terms of a ratio of successful single photon emission and emission from a background noise. On the fundamental side, they uncover manifestation of nonclassicality of light in different settings of interference experiments. On the technical side, they allow direct comparison of single photon sources beyond the HBT measurement. Experimental layouts are simple extensions of existing experiments, therefore these criteria can be immediately implemented in laboratories. Two-copy criteria can be also adapted to recent time multiplexed multiphoton sources \cite{lodahlSP, senellart, goetzinger}. Also the criteria can investigate nonclassicality of multiphoton light \citep{luo,weihs2,genovese}.

Last but not least, we present a methodology which can be straightforwardly extended to find other new nonclassicality criteria based on different settings and more copies \citep{weigh}. It will allow a better understanding of even small impacts of multiphoton contributions in single photon states and their influence in linear optical protocols \citep{innocenti} of quantum technology. Such a catalog of detection layouts for different variable nonclassicality criteria can be efficiently implemented using current state of the art integrated optical technology. It will be then applicable as versatile detector for quantum technology with single photon states.

\section{Acknowledgement}

We thank Miroslav Je\v zek and Darren W. Moore for a fruitful discussion. This work was supported by the Czech
Science Foundation (17-26143S), national funding from the MEYS, funding from European Union’s Horizon 2020 (2014-2020) under grant agreement No 731473 (QuantERA project HYPER-U-P-S No 8C18002) and also by IGA-
Prf-2018-010 and IGA-Prf-2019-010.

\section{Appendix:Propagation of classical states through interfering layouts}
The propagation of $n$ incoming coherent states through an interfering network is determined by a matrix $A$ that transforms the vector of amplitudes $v=(\alpha_1,...,\alpha_n)$ to a new vector $u=(\alpha'_1,...,\alpha'_n)$ by the relation
\begin{equation}
u=A v.
\label{trans}
\end{equation}
A matrix corresponding to a beam splitter holds
\begin{equation}
A_{BS}=
  \begin{bmatrix}
    \sqrt{T} & \sqrt{1-T} \\
    -\sqrt{1-T} & \sqrt{T}
  \end{bmatrix},
\end{equation}
where $T$ is the transmission of the beam splitter. The impact of a Mach-Zehnder interferometer is given by
\begin{equation}
A_{\otimes 1}=
  \begin{bmatrix}
    \sqrt{T_1 T_2} & \sqrt{R_1 R_2} \\
   \sqrt{R_1 R_2}e^{i \phi} & \sqrt{T_1 T_2}e^{i \phi}
  \end{bmatrix},
\end{equation}
where $T_{1,2}$ ($R_{1,2}$) is transmission (reflection) of $\mbox{BS}_{1,2}$ depicted in Fig.~ \ref{setups} a) and $\phi$ is a relative phase acquired between two paths by which the light propagates towards detectors. The outputs of  two-copies interfering layouts including Hong-Ou-Mandel interferometer and layout shown in Fig.~\ref{setups} b) are determined from
\begin{equation}
A_{hom,\otimes 2}=\begin{bmatrix}
    \sqrt{T_1} & \sqrt{R_1} & 0 \\
   -\sqrt{R_1 T_2} & \sqrt{T_1 T_2} & \sqrt{1-T_2}\\
   \sqrt{R_1 R_2} & -\sqrt{T_1 R_2} & \sqrt{T_2}
  \end{bmatrix}.
\end{equation}
Again, $T_{1,2}$ and $R_{1,2}$ correspond to transmission and reflection of the the relevant beam splitters. The layouts differ themselves by input vectors. Whereas the input state is $v=(0,\alpha_1,\alpha_2)$ in case of Hong-Ou-Mandel interferometer, the incoming state of the interfering network in Fig.~\ref{setups} b) is $v=(\alpha_1,\alpha_2,0)$.

Any click statistics can be expressed by no click probabilities. Let $\mathbf{K}$ denotes a group of detectors for which $P_{0,\mathbf{K}}$ quantifies a probability that no detector in $\mathbf{K}$ register a click. For coherent states, the probability  yields
\begin{equation}
P_{0,\mathbf{K}}=\Pi_{i\in \mathbf{K}}\exp(-\vert \alpha_i \vert ^2),
\end{equation}
where $i$ is an index of a detector in $\mathbf{K}$ and $\alpha_i$ is an amplitude of an output coherent state which is measured by the detector. According to (\ref{trans}), each amplitude $\alpha_i$ depends linearly on the amplitudes of incoming coherent states. Since the phases of these coherent states are randomized, the probabilities $P_{0,\mathbf{K}}$ have to be integrated over phase of each input state
\begin{equation}
\bar{P}_{0,\mathbf{K}}=\int_{\phi_1}...\int_{\phi_n}P_{0,\mathbf{K}}(\phi_1,...,\phi_n)\mathrm{d}\phi_1...\mathrm{d}\phi_n,
\end{equation}
where $\phi_i$ represents a phase of the incoming coherent states in the mode $i$, i. e. $\alpha_i=\vert \alpha_i \vert e^{i\phi_i}$. It consequently means the probabilities $\bar{P}_{0,\mathbf{K}}$ are expressed by Bessel functions. The success and error probabilities are expressed as linear combination of $\bar{P}_{0,\mathbf{K}}$, particularly
\begin{eqnarray}
P_{s,BS}&=&1-\bar{P}_{0,(1)}\\ \nonumber
P_{e,BS}&=&1-\bar{P}_{0,(1)}-\bar{P}_{0,(2)}+\bar{P}_{0,(1,2)}\\ \nonumber
P_{s,\otimes 1}&=&1-\bar{P}_{0,(1)}\\ \nonumber
P_{e,\otimes 1}&=&1-\bar{P}_{0,(1)}-\bar{P}_{0,(2)}+\bar{P}_{0,(1,2)}\\ \nonumber
P_{s,hom}&=&1-\bar{P}_{0,(1)}-\bar{P}_{0,(2)}+\bar{P}_{0,(1,2)}\\ \nonumber
P_{e,hom}&=&1-\bar{P}_{0,(1)}-\bar{P}_{0,(2)}-\bar{P}_{0,(3)}+\bar{P}_{0,(1,2)}\\ \nonumber
&+&\bar{P}_{0,(2,3)}+\bar{P}_{0,(1,3)}-\bar{P}_{0,(1,2,3)} \\
\nonumber
P_{s,\otimes 2}&=&1-\bar{P}_{0,(1)}-\bar{P}_{0,(2)}+\bar{P}_{0,(1,2)}\\ \nonumber
P_{e,\otimes 2}&=&1-\bar{P}_{0,(1)}-\bar{P}_{0,(2)}-\bar{P}_{0,(3)}+\bar{P}_{0,(1,2)}\\ \nonumber
&+&\bar{P}_{0,(2,3)}+\bar{P}_{0,(1,3)}-\bar{P}_{0,(1,2,3)},
\end{eqnarray}
where subscripts BS, $\otimes 1$, hom and $\otimes 2$ identify the layout which is employed for the detection of success and error probabilities $P_s$, $P_e$.


\begin{thebibliography}{99}
\bibitem{Eisman}
Eisaman M D, Fan J, Migdall A and Polyakov S V 2011 Invited review article: single-photon sources and detectors. Rev. Sci. Instrum. \textbf{82} 071101
\bibitem{Thew}
Gisin N and Thew R 2007 Quantum communication Nat. Photonics \textbf{1} 165–171
\bibitem{BB84}
Bennett C and Brassard G 1984 Quantum cryptography: Public key distribution and coin tossing, in Proceedings of IEEE International Conference on Computers, Systems, and Signal Processing, Bangalore, India (IEEE, New York, 1984), pp. 175–179
\bibitem{Kimble}
Kimble H J 2008 The quantum internet Nat. \textbf{453} 1023-1030
\bibitem{simulation}
Aspuru-Guzik A, Walther P 2012 Photonic quantum simulators Nat. Phys. \textbf{8} 285-291
\bibitem{Lodahl}
Lodahl P 2018 Quantum-dot based photonic quantum networks Quantum Sci. Technol. \textbf{3} 013001
\bibitem{mandel}
Kimble H J, Dagenais M and Mandel L 1977 Photon Antibunching in Resonance Fluorescence Phys. Rev. Lett. \textbf{39} 691
\bibitem{grangier}
Grangier P, Roger G and Aspect A 1986 Experimental Evidence for a Photon Anticorrelation Effect on a Beam Splitter: A New Light on Single-Photon Interferences, Eur. Phys. Lett. \textbf{1} 173
(1986).
\bibitem{Ekert}
Ekert A K, Quantum cryptography based on Bell’s theorem 1991 Phys. Rev. Lett. \textbf{67} 661
\bibitem{Dusek}
Scarani V, Bechmann-Pasquinucci H, Cerf N J, Du\v sek M, L\" utkenhaus N and Peev M 2009 The security of practical quantum key distribution, Rev. Mod. Phys. \textbf{81} 1301
\bibitem{Burnett}
Holland M J and Burnett K 1993 Interferometric detection of optical phase shifts at the Heisenberg limit Phys. Rev. Lett. \textbf{71} 1355
\bibitem{singlePhoton}
Michler P, Kiraz A, Becher C, Schoenfeld W V, Petroff P M, Zhang L, Hu E and Imamoglu A 2000 A quantum dot single-photon turnstile device Science \textbf{290} 2282
\bibitem{lodahlC}
Arcari M, S\" ollner I, Javadi A, Lindskov Hansen S, Mahmoodian S, Liu J, Thyrrestrup H, Lee E J, Song J D, Stobbe S and Lodahl P 2014 Near-Unity Coupling Efficiency of a Quantum Emitter to a Photonic Crystal Waveguide
Phys. Rev. Lett. \textbf{113} 093603
\bibitem{lodahlSP}
Kir\v sansk\. e G, Thyrrestrup H, Daveau R S, Dreessen C L, Pregnolato T, Midolo L, Tighineanu P, Javadi A, Stobbe S, Schott R, Ludwig A, Wieck A D, In Park S, Song J D, Kuhlmann A V, S\" ollner I, L\" obl M C, Warburton R J and Lodahl P 2017 Indistinguishable and efficient single photons from a quantum dot in a planar nanobeam waveguide
Phys. Rev. B \textbf{96}, 165306
\bibitem{senellart}
Somaschi N, Giesz V, De Santis L, Loredo J C, Almeida M P, Hornecker G, Portalupi S L, Grange T, Ant\' on C, Demory J, G\' omez C, Sagnes I, Lanzillotti-Kimura N D, Lema\' itre A, Auffeves A, White A G, Lanco L and Senellart P 2016 Near-optimal single-photon sources in the solid state, Nat. Phot. \textbf{10} 340–345
\bibitem{goetzinger}
Chu X-L, G\" otzinger S and Sandoghdar V 2017 A single molecule as a high-fidelity photon gun for producing intensity-squeezed light Nat. Phot. \textbf{11} 58–62
\bibitem{revNoncl}
Miranowicz A, Bartkowiak M, Wang X, Liu Y-X and Nori F 2010 Testing nonclassicality in multimode fields: A unified derivation of classical inequalities Phys. Rev. A \textbf{82} 013824
\bibitem{Agarwal}
Sperling J, Vogel W, and Agarwal G S 2012 Sub-Binomial Light Phys. Rev. Lett. \textbf{109} 093601
\bibitem{Walmsley}
Rigovacca L, Di Franco C, Metcalf B J, Walmsley I A and Kim M S 2016 Nonclassicality Criteria in Multiport Interferometry Phys. Rev. Lett. \textbf{117} 213602
\bibitem{Sperling}
Sperling J, Eckstein A, Clements W R, Moore M, Renema J J, Kolthammer wW S, Nam S W, Lita A, Gerrits T, Walmsley I A, Agarwal G S and Vogel W 2017 Identification of nonclassical properties of light with multiplexing layouts Phys. Rev. A \textbf{96} 013804
\bibitem{filipLach}
Filip R and Lachman L 2013 Hierarchy of feasible nonclassicality criteria for sources of photons Phys. Rev. A \textbf{88} 043827
\bibitem{nonG1}
Je\v zek M, Straka I, Mi\v cuda M, Du\v sek M, Fiur\' a\v sek J and R. Filip 2011 Experimental Test of the Quantum Non-Gaussian Character of a Heralded Single-Photon State Phys. Rev. Lett. \textbf{107} 213602
\bibitem{nonGN}
Straka I, Lachman L, Hlou\v sek J, Mikov\' a M, Mi\v cuda M, Je\v zek M, and Filip R 2018 Quantum non-Gaussian multiphoton light npj Quantum Information \textbf{4} 4 
\bibitem{Obsil}
Ob\v sil P, Lachman L, Pham T, Le\v sund\' ak A, Hucl V, \' C\v i\v zek M, Hrabina J, \v C\' ip O, Slodi\v cka L and R. Filip 2018 Nonclassical Light from Large Ensembles of Trapped Ions Phys. Rev. Lett. \textbf{120} 253602
\bibitem{tuning1}
Midolo L, Hansen S L, Zhang W, Papon C, Schott R, Ludwig A, Wieck A D, Lodah P and Stobbe S 2017 Electro-optic routing of photons from single quantum dots in photonic integrated circuits  Opt. Express \textbf{25} 33514-33526
\bibitem{tuning2}
Flamini F, Magrini L, Rab A S, Spagnolo N, D'Ambrosio V, Mataloni P, Sciarrino F, Zandrini T, Crespi A, Ramponi R and Osellame R 2015 Thermally reconfigurable quantum photonic circuits at telecom wavelength by femtosecond laser micromachining, Light: Science \& Applications \textbf{4} e354
\bibitem{integratedOptics}
O'Brien J L, Furusawa A, Vu\v ckovi\' c J 2009 Photonic quantum technologies Nat. Phot. \textbf{3} 687–695
\bibitem{integrateOptics2}
Meany T, Grafe M, Heilmann R, Perez-Leija A,
Gross S, Steel M J, Withford M J and Szameit A 2015 Laser written circuits for quantum photonics Laser Photon.
Rev. \textbf{9} 1863.
\bibitem{integrateOptics3}
Carolan J, Harrold C, Sparrow C, Martin-Lopez E, Russell N J, Silverstone J W, Shadbolt P J, Matsuda N, Oguma M, Itoh M, Marshall G D, Thompson M G, Matthews J C F, Hashimoto T, O’Brien J L and Laing A 2015 QUANTUM OPTICS. Universal linear optics Science \textbf{349} 711
\bibitem{glauber}
Glauber R J 1963 Coherent and Incoherent States of the Radiation Field Phys. Rev. \textbf{131} 2766
\bibitem{hong}
Hong C K, Ou Z Y and Mandel L 1987 Measurement of subpicosecond time intervals between two photons by interference Phys. Rev. Lett. \textbf{59} 2044
\bibitem{hongCr}
Mandel L 1983 Photon interference and correlation effects produced by independent quantum sources Phys. Rev. A \textbf{28} 929
\bibitem{luo}
Qi L, Manceau M, Cavanna A, Gumpert F, Carbone L, de Vittorio M, Bramati A, Giacobino E, Lachman L, Filip R and Chekhova M 2018 Multiphoton nonclassical light from clusters of single-photon emitters
NJP \textbf{20} (7) 073013
\bibitem{weihs2}
Khoshnegar M, Huber T, Predojevi\' c A, Dalacu D, Prilm\" uller M, Lapointe J, Wu X, Tamarat P, Lounis B, Poole P, Weihs G and Majedi H 2017 A solid state source of photon triplets based on quantum dot molecules
Nat. Comm. \textbf{8} 15716
\bibitem{genovese}
Moreva E, Traina P, Forneris J, Degiovanni I P, Ditalia Tchernij S, Picollo F, Brida G, Olivero P and Genovese M 2017 Direct experimental observation of nonclassicality in ensembles of single-photon emitters
Phys. Rev. B \textbf{96} 195209
\bibitem{weigh}
Agne S, Kauten T, Jin J, Meyer-Scott E, Salvail J Z, Hamel D R, Resch K J, Weihs G and Jennewein T 2017 Observation of Genuine Three-Photon Interference Phys. Rev. Lett. \textbf{118} 153602
\bibitem{innocenti}
Giordani T, Polino E, Emiliani S, Suprano A, Innocenti L, Majury H, Marrucci L, Paternostro M, Ferraro A, Spagnolo N and Sciarrino F 2019 Experimental Engineering of Arbitrary Qudit States with Discrete-Time Quantum Walks Phys. Rev. Lett. \textbf{122} 020503


\end{thebibliography}
\end{document}